\documentclass{appolb}
\usepackage{epsfig}
\newcommand{\be}{\begin{equation}}
\newcommand{\ee}{\end{equation}}
\newcommand{\ba}{\begin{eqnarray}}
\newcommand{\ea}{\end{eqnarray}}
\newcommand{\bc}{\begin{center}}
\newcommand{\ec}{\end{center}}
\newcommand{\bfig}{\begin{figure}}
\newcommand{\efig}{\end{figure}}
\newcommand{\f}[2]{\frac{#1}{#2}}
\newcommand{\g}{\gamma}

\newcommand{\al}{\alpha}

\newcommand{\rr}[4]{#1, {\it #2 \/}{\bf #3} #4}

\begin{document}
\title{Testing resummed NLO-BFKL kernels\thanks{To appear in the special issue 
of 
Acta Physica Polonica to celebrate the 65th Birthday of Professor Jan 
Kwiecinski}
}
\author{R. Peschanski
\address{
CEA/DSM/SPhT, Unit\'e de recherche 
associ\'ee 
au CNRS, \\
CE-Saclay, F-91191 Gif-sur-Yvette Cedex, France\thanks{
	E-mail: {pesch@spht.saclay.cea.fr}}
}}
\maketitle
\begin{abstract}
We propose a new method to test the (resummed) next-to-leading-order BFKL  
evolution kernels  
 using the Mellin transformed $j$-moments of  the proton 
structure function $F_2.$ 
\end{abstract}

\section{How to test BFKL evolution equations?}

The Balitsky Fadin Kuraev Lipatov (BFKL) evolution equation \cite{bfkl}, 
derived 
in the framework of 
perturbative QCD,  has held the attention
of the scientific community since a long time. The summation of leading 
logarithms of 
energy in the perturbative expansion gives valuable tools for the investigation 
of deep-inelastic scattering at small $x_{Bj}$ (equivalently large energy 
squared $W^2 \sim Q^2/x_{Bj}$). Indeed, the first experimental results from 
HERA 
confirmed the existence of a strong rise of the proton structure function $F_2$ 
with energy in agreement with the trends implied by the solution of the BFKL 
equation. It has 
been 
possible \cite{old} to describe the old data at small $x_{Bj}$ and  in a certain 
range of $Q^2.$ However the price to pay was  to get a  phenomenological value 
of the 
intercept (the exponent 
of $1/x_{Bj}$ in the BFKL formula, see later in the text)   
less than the 
predicted range (the corresponding value of the  strong coupling constant  is 
$\al _s \sim .1$ 
instead of 
$\sim .2$). This was revealing the need for  rather sizable higher 
order corrections.

At the next-to-leading log level, these  corrections have been 
calculated after much efforts \cite{next} and appeared to be so large that they 
overshoot the expected phenomelogical effect and could even invalidate the whole 
approach. 
Soon after, it was realized \cite{salam} 
that the main problem came from the existence of spurious singularities which 
ought to be cancelled 
by an appropriate resummation at all orders of the perturbative expansion. This 
is required  by the QCD renormalization group. Indeed, various resummation 
schemes have been proposed \cite{salam,autres,lipatov} which satisfy the 
renormalization group 
requirements while retaining the exact value of the 
next-leading log term in 
the 
BFKL kernel computed in Refs.\cite{next}. Hence, the constraints can be  
satisfied and the next-to-leading 
order introduced without destroying the whole scheme. 

However, the 
resummation schemes  possess some ambiguity, since 
higher order logs (beyond the next-to-leading ones) are not known. Such 
variations appear, {\it e.g.} in Ref.\cite{salam}, where four different 
resummation schemes have been proposed. These schemes, denoted {\it scheme 
1,...4} in the following, will be the subject of our present study. Other 
schemes have been 
proposed 
\cite{autres,lipatov} and will be studied as well later \cite{us}. It is worth 
to confront these various schemes with data, 
to check their validity and distinguish between different resummation options.

Precise phenomenological tests of QCD evolution equations are one of the main 
goals of deep inelastic scattering phenomenology. For DGLAP evolution 
\cite{dglap}, it has 
been possible 
to test it in various ways with next-to-leading log $Q^2$ (NLO) corrections and 
it works quite well in a large range of $Q^2.$ 
Testing precisely BFKL 
evolution beyond leading order is much more difficult. The main problem is the 
complicated 
 mismatch between QCD perturbative and non perturbative  inputs, since the 
corresponding factorization properties, {\it i.e.  $k_T$ factorization} 
\cite{kt},  are 
more involved than for DGLAP evolution. A 
way out
could be to stay within the perturbative regime by using only massive or highly 
virtual colliding particles, like 
$\g^*\!-\!\g^*$ 
scattering, but the data are yet too imprecise that no definite conclusion can 
be 
drawn. Note also that  some perturbative QCD ingredients (such as the so-called 
``impact factors'') are not yet but will be soon available
\cite{bartels}.

In the present paper, we propose a method for  
testing   the (resummed) BFKL 
predictions
for the proton structure functions, {\it via} a  transformation to Mellin space. 
 
On the one hand,  the present set of data allows for a precise determination of 
the   
Mellin transform of $F_2$  in 
a 
large range of $Q^2$ and $j,$ the Mellin 
conjugate of $x_{Bj},$  considered as a 
continuous variable. On the other hand,  the BFKL predictions at leading order 
and beyond 
are 
easier to formulate in Mellin space since one can obtain  tests of the 
evolution kernels which are essentially dependent on the calculable 
perturbative part. In our formulation, we assume that $k_T$ factorization 
remains valid for the resummed kernel. An improved 
formulation of  $k_T$ factorization containing the  NLO $\g^*$ impact factors 
when 
they will be available, will allow to refine our study in the future.

The proposed method  has the following features. It treats in parallel both  LO 
and (resummed) NLO BFKL kernels. It uses   a
$k_T$ factorized formulation  of the structure functions which includes the 
factorized Green function solution derived in Ref. \cite{autres}. The present 
essay is of introductory nature, and present   a first, non sophisticated,
phenomenological  investigation of the proton 
structure functions \cite{salomez,salomez'}, where we  compare the LO-DGLAP
 GRV parametrization \cite{grv} of proton structure functions with the BFKL 
kernel predictions. In a 
forthcoming publication \cite{us}, we shall give an  extensive and more 
systematic study 
using our method.

The plan of contents is organized as follows; In the next section, we start by 
expressing the  LO BFKL predictions  in Mellin 
 space, defining three characteristic relations. In section {\bf 3} we elaborate 
the corresponding 
set of predictions for the resummed NLO-BFKL kernels. An   application to the 
DGLAP/BFKL 
comparison is presented in section {\bf 4}. Section {\bf 5} provides a 
conclusion and   an
outlook on future work.

\section{BFKL predictions in Mellin space}

The formulation of the proton structure functions in the (LO) BFKL 
approximation 
can be expressed as follows \cite{old}:

\begin{equation}
\pmatrix{F_T \cr F_L \cr G}=\int 
\frac{d\gamma}{2i\pi}\left(\frac{Q^2}{Q_0^2}\right)^{\gamma} 
e^{\frac {\al_s N_c}{\pi} \chi_{L0}(\gamma) \ln \left({1}/{x_{Bj}}\right)}
\pmatrix{h_T \cr h_L \cr 
1}\omega(\gamma)
\label{bfkl}
\end{equation}
where one has written the BFKL kernel as
\begin{equation}
\chi_{L0}(\gamma)=2\psi(1)-\psi(\gamma)-\psi(1-\gamma)\ .
\label{kernel}
\end{equation}
In formula (\ref{bfkl}), with conventional notations, $F_T , F_L ,  G$ stand 
respectively for transverse , longitudinal and gluon structure functions, 
$\al_s$ is the (fixed) coupling constant,
$\omega(\gamma)$ is an (unknown) non-perturbative coupling to the proton while 
\begin{equation}
\pmatrix{h_T \cr 
h_L}=\frac{\al_s}{3\pi\gamma}\frac{(\Gamma(1-\gamma)\Gamma(1+\gamma))^3}{\Gamma
(
2-2\gamma)\Gamma(2+2\gamma)}\frac{1}{1-\frac{2}{3}\gamma}\pmatrix{(1+\gamma)
(1-
\frac{\gamma}{2}) \cr \gamma(1-\gamma)}\ ,
\end{equation}
correspond to the known perturbative couplings to the photon, usually called LO 
``impact factors'' in the literature. Note that, in the framework of $k_T$ 
factorization,  $\gamma$ plays the r\^ole of a ``running'' 
anomalous dimension, whose physical value is determined by the integration of  
(\ref{bfkl}). As already 
mentionned in the introduction,  formula (\ref{bfkl}) gives rise to an 
interesting {\it effective}  BFKL 
phenomenology \cite{old} in the small $x_{Bj}$ region, but it has to rely on 
the 
parametrization of the unknown non perturbative function $\omega(\gamma)$ and 
leads to  values of $\al_s$ quite smaller than expected.

Mellin-transforming (\ref{bfkl}) in $j$-space, one easily finds
\begin{equation}
\pmatrix{\tilde F_T \cr\tilde F_L \cr\tilde G}=\int 
\frac{d\gamma}{2i\pi}\left(\frac{Q^2}{Q_0^2}\right)^{\gamma} 
\frac{1}{j-1-\frac{\al_s N_c}{\pi}\chi_{L0}(\gamma)}\pmatrix{h_T \cr h_L \cr 
1}\omega(\gamma)\ .
\label{mellin}
\end{equation}
Looking for the poles in $\gamma,$ it is straightforward to use the residue 
formula\footnote{Care is to be taken of the contour at infinity, see the second 
reference of \cite{us}.} and get
\begin{equation}
\!\pmatrix{\tilde F_T \cr\tilde F_L \cr\tilde G}=\sum_{i} 
\frac{1}{\frac{\al_s 
N_c}{\pi}[-\chi_{L0}'(\gamma_i(j))]}\left(\frac{Q^2}{Q_0^2}\right)^{\gamma_i(j)
}
\!\pmatrix{h_T(\gamma_i(j)) 
\cr h_L(\gamma_i(j)) \cr 1}\ \omega(\gamma_i(j))
\label{poles}
\end{equation}
where  $\gamma_i(j)$ are the ($\gamma_i < 1/2 $) roots of the equation
\begin{equation}
j-1=\frac{\al_s N_c}{\pi}\chi_{L0}(\gamma_i(j))\ .
\label{roots}
\end{equation}
In fact, with good accuracy at large enough $Q^2,$ (comparable to the leading 
twist approximation in 
DGLAP evolution), one can only retain the rightmost  pole $\gamma_1(j).$  We 
are 
thus left with the following simple formula as a starting point of our analysis:
\begin{equation}
\pmatrix{\tilde F_T \cr\tilde F_L \cr\tilde G}\approx 
\frac{1}{\frac{\al_s 
N_c}{\pi}[-\chi_{L0}'(\gamma_1(j))]}\left(\frac{Q^2}{Q_0^2}\right)^{\gamma_1(j)
}\pmatrix{h_T(\gamma_1(j)) \cr 
h_L(\gamma_1(j)) \cr 1}\ \omega(\gamma_1(j))\ .
\label{final}
\end{equation}

From equation (\ref{final}), three model-independent predictions, {\it i.e.} 
independent of non-perturbative assumptions,  can be drawn:

{\bf i)}  The Mellin transform of $F_2 \equiv F_T+F_L$ should verify:
\be
\ln \tilde F_2 (j,{Q^2}) = \gamma_1(j) \ln ({Q^2}) + f(j)
\label{i)}
\ee
in some range of $j$ near $1$ where the BFKL equation is expected to be 
relevant. The function $f(j)$ regroups all $Q^2$-independent terms in 
(\ref{final}).

{\bf ii)} $\gamma_1(j),$ extracted from  (\ref{final}) as the slope in $\ln 
({Q^2}),$ should verify the equation
(\ref{roots}) for the anomalous dimension, namely
\be
\chi_{L0}(\gamma_1(j))\equiv \frac{\pi}{\al_s 
N_c} \ (j\!-\!1)\ ,
\label{ii)}
\ee
with constant ${\al_s },$  and $\chi_{L0}$ given by (\ref{kernel}).

{\bf iii)} The gluon structure function (one may also choose the obervable 
$F_L$) should verify, {\it via} Mellin transform:
\begin{equation}
\ln(\tilde{G}(j,Q^2))=\ln\left(\tilde{F_2}(j,Q^2)\right)-\gamma_1(j)\ 
\ln\left(h_T+
h_L\right)\ .
\label{iii)}
\end{equation}
The predictions (\ref{i)}), (\ref{ii)}) (\ref{iii)}) represents a stringent set 
of constraints which have to be verified by the Mellin-transformed of the 
proton structure functions in a region $j$ near 1. In fact we will confirm that 
the (LO) BFKL kernel 
does not pass this step.

\section{Resummed NLO-BFKL predictions in Mellin space}

Interestingly, using a reasonable $k_T$ factorized ansatz\footnote{$k_T$ 
factorization has not been yet proven at NLO, but is a reasonable ansatz 
fulfilling the known theoretical requirements on the kernel properties discussed 
in \cite{autres}.}, the predictions {\bf i)}-{\bf iii)} remain valid for 
NLO-BFKL resummation kernels, up to specific modifications due to the running of 
the coupling constant. 

 Let us formulate the (resummed) NLO-BFKL structure functions in 
Mellin space as follows:
\begin{equation}
\pmatrix{\tilde F_T \cr\tilde F_L \cr\tilde G}=\int 
\frac{d\gamma}{2i\pi}\left(\frac{Q^2}{\Lambda_{QCD}^2}\right)^{\gamma}
e^{-\f {1}{b(j\!-\!1)}\ X(\g,j)}
\pmatrix{h_T \cr h_L \cr 
h_G}\eta(\gamma)\ ,
\label{mellin-NLO}
\end{equation}
where, by definition
\be
{\partial \over \partial \g} X(\g,j)\equiv \chi_{NLO}(\g,j)\ .
\label{X}
\ee
The function $X(\g,j)$  appears in the solution
of  the Green function derived\footnote{The second variable of $X(\g,j)$ in 
(\ref{X}) corresponds to the choice of a reference scale $\mu\to j-1$ dictated 
by the treatment of  the Green function fluctuations near the  saddle-point 
\cite{autres}.} from the 
renormalization-group improved small-$x_{Bj}$ equation \cite{autres}, 
$\chi_{NLO}(\g,j)$ is the resummed NLO-BFKL kernel and $b=11-2/3\ N_f/N_c$  
defines 
the running of the coupling constant 
\be
\f{N_c}{\pi}\ \al_s(Q^2)=\f{1}{b\ln\left(Q^2/\Lambda_{QCD}^2\right)}\ .
\label{al}
\ee

Before going further, let us comment  formula (\ref{mellin-NLO}). This 
formula captures the (large) $Q^2$-dependent part of the gluon Green function 
which has 
been shown to have
factorization properties \cite{autres}. In fact the non perturbative 
contribution has been factorized out in the  function $\eta(\g).$ Some 
unknown $Q^2$-dependence  may still remain in the NLO 
contributions to the impact factor 
vector $(h),$ which we  neglected in the present analysis.

Starting from this ansatz, let us derive the NLO constraints similar to 
(\ref{i)})-(\ref{iii)}).
At large enough $Q^2/\Lambda_{QCD}^2,$ one can use the saddle-point appoximation 
to evaluate  (\ref{mellin-NLO}). Assuming that the perturbative impact factors 
 and the non-perturbative function $\eta$ do not vary much\footnote{We do not 
take into account modifications  {\it e.g.} coming from  powers of $\g$ in the 
prefactors which 
may shift the saddle point \cite{autres}. We thus assume a smoothness of the 
structure function integrand around the saddle-point in agreement with the 
phenomenology \cite{us}.},  the 
saddle-point condition reads
 \be
j\!-\!1  \sim\f {1}{b\ \ln\left(Q^2/\Lambda_{QCD}^2\right)}\ 
\chi_{NLO}(\bar\g,j) =
\f{N_c\ \al_s(Q^2)}{\pi} \ \chi_{NLO}(\bar\g,j)\ ,
 \label{saddle}
 \ee
where $\bar\g(j,Q^2)$ is the saddle-point value. Relation (\ref{saddle}) is 
nothing else than the  NLO extension of condition {\bf ii)} of 
(\ref{ii)}) to the case of a running coupling constant (\ref{al}).  

Inserting 
the saddle 
point defined by (\ref{saddle}) in formula (\ref{mellin-NLO}), one obtains the 
new set of constraints at NLO level as follows:

{\bf i)}  The Mellin transform of $F_2 \equiv F_T+F_L$ verifies:
\be
{\partial \over \partial \ln ({Q^2})} \ln \tilde F_2 (j,{Q^2}) \ \sim \ \bar 
\g(j,Q^2) 
\label{i+)}
\ee
where $\bar \g(j,Q^2)$ is now a smoothly $Q^2$-dependent effective anomalous 
dimension defined by the following  property.

{\bf ii)} $\bar \g(j,Q^2)$  verifies the anomalous dimension equation, namely
\be
\chi_{NLO}(\bar \g(j,Q^2))\equiv \frac{\pi}{\al_s (Q^2)
N_c} \ (j\!-\!1)\ ,
\label{ii+)}
\ee
where $\chi_{NL0}$ is one of the  resummed NLO-BFKL  candidate kernels
proposed in the 
literature.

{\bf iii)} The gluon structure function (one may also choose the obervable 
$F_L$)  verifies, {\it via} Mellin transform:
\begin{equation}
\ln(\tilde{G}(j,Q^2))=\ln\left(\tilde{F_2}(j,Q^2)\right)-\bar \g(j,Q^2)\ 
\ln\left[h_T(\bar \g)+
h_L(\bar \g)\right]\ ,
\label{iii+)}
\end{equation}
where NLO effects of impact factors are neglected.

The interest of the relations (\ref{i+)})  (\ref{ii+)}) (\ref{iii+)}) is that 
they are formally similar with the LO ones by the direct substitution of the LO 
kernel by the NLO ones and of a fixed coupling constant by the running one at 
one-loop. They are only approximate, since they rely on a saddle-point 
approximation which may not be always justified (see \cite{autres} for a 
discussion). However, in the present context, the validity of the saddle-point 
approximation can be 
tested directly from the phenomenological analysis. Due to the observed 
smoothness of the Mellin transforms,  we do not expect large corrections to the 
saddle-point results.

\section{Application: the ``proximity'' between DGLAP and NLO-BFKL}

In this section, we want to check the reliability of the Mellin space  method by 
a study of DGLAP parametrizations of the data \cite{salomez} . It is well-known 
that DGLAP 
parametrizations fit well the data in a large range of $x_{Bj}$ and $Q^2 >1\ 
GeV.$ Choosing such a parametrization of structure functions,  namely the   GRV 
set 
of structure functions \cite{grv}, we are able to Mellin transform them easily, 
and thus discuss the comparison between DGLAP and LO/NLO BFKL evolution 
equations. DGLAP evolution is automatically obeyed by the input functions and we 
want to compare them with BFKL evolution using relations (\ref{i)})  (\ref{ii)}) 
(\ref{iii)}) for LO and (\ref{i+)})  (\ref{ii+)}) (\ref{iii+)}) for NLO. The 
physical question we ask in this application is whether or not there may exist a 
compatibility between DGLAP and BFKL evolution equations.

\begin{figure}[htb]
\begin{center}
\includegraphics[width=5cm]{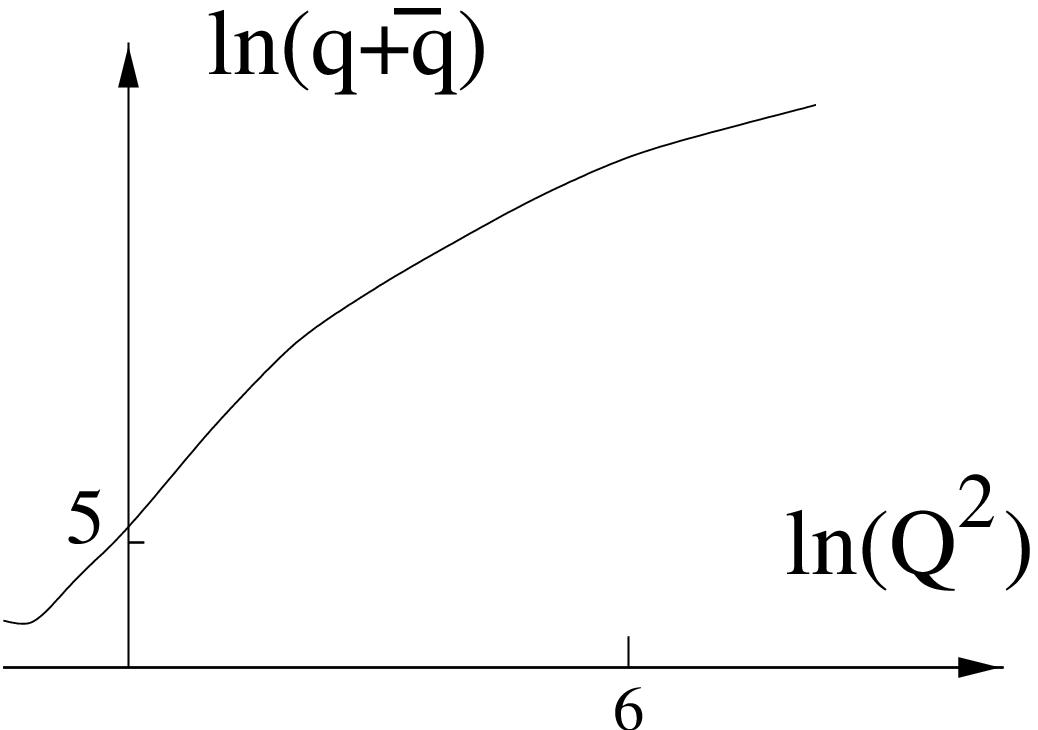} \hspace {2cm}
\includegraphics[width=5cm]{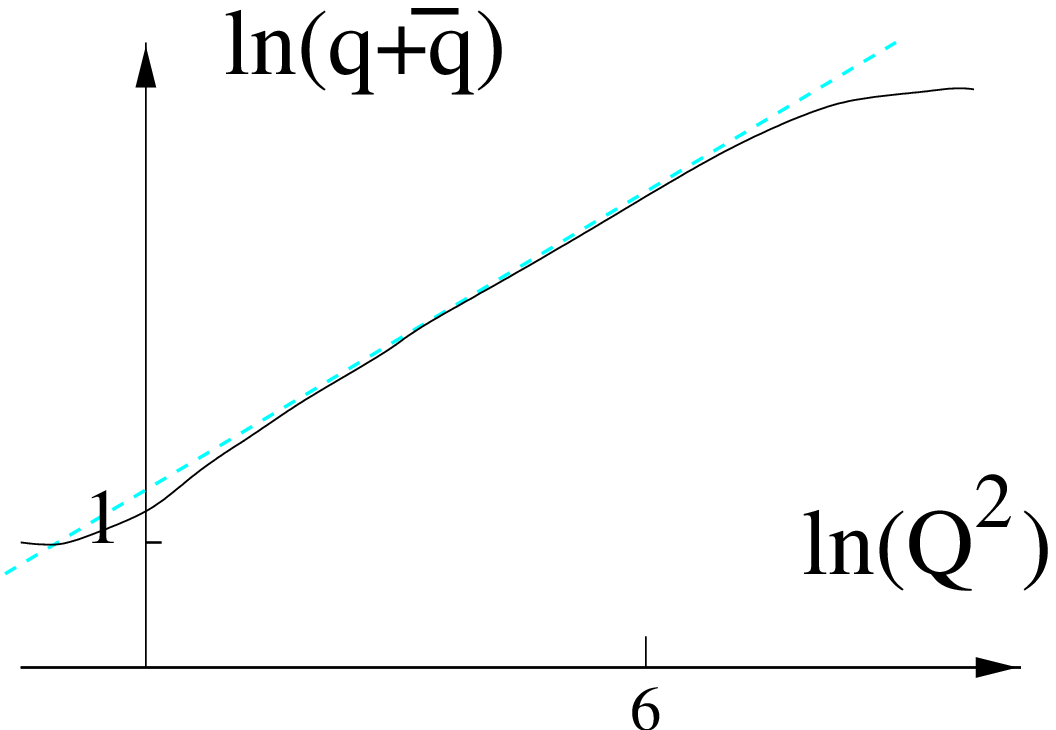}
\includegraphics[width=5cm]{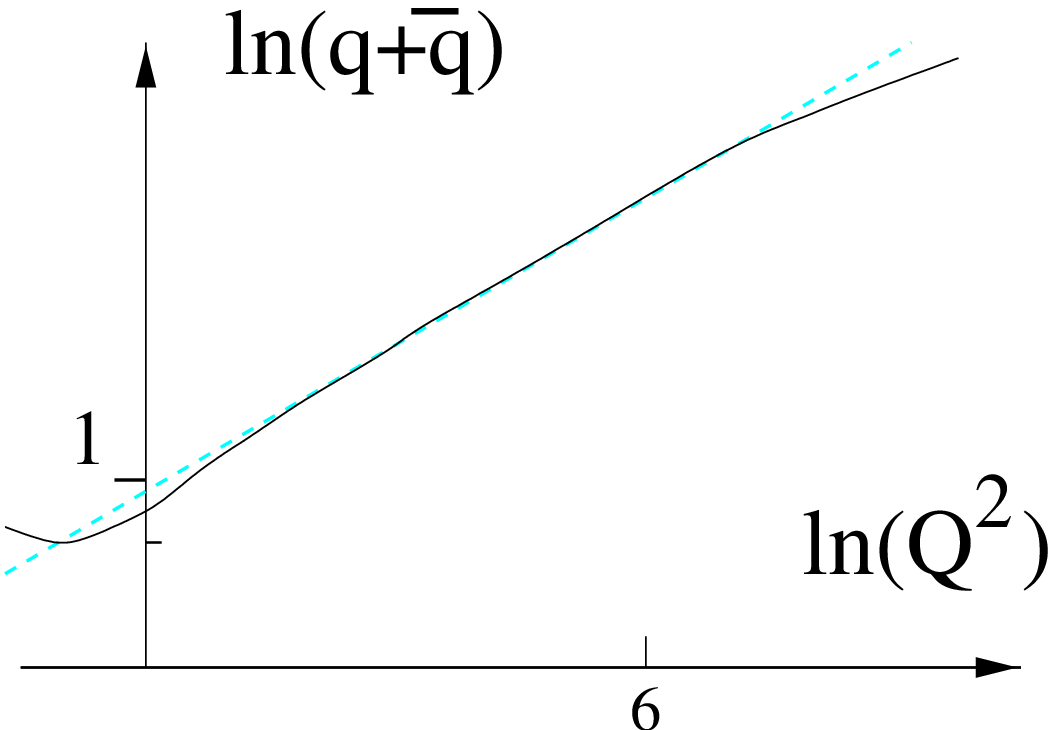} \hspace {2cm}
\includegraphics[width=5cm]{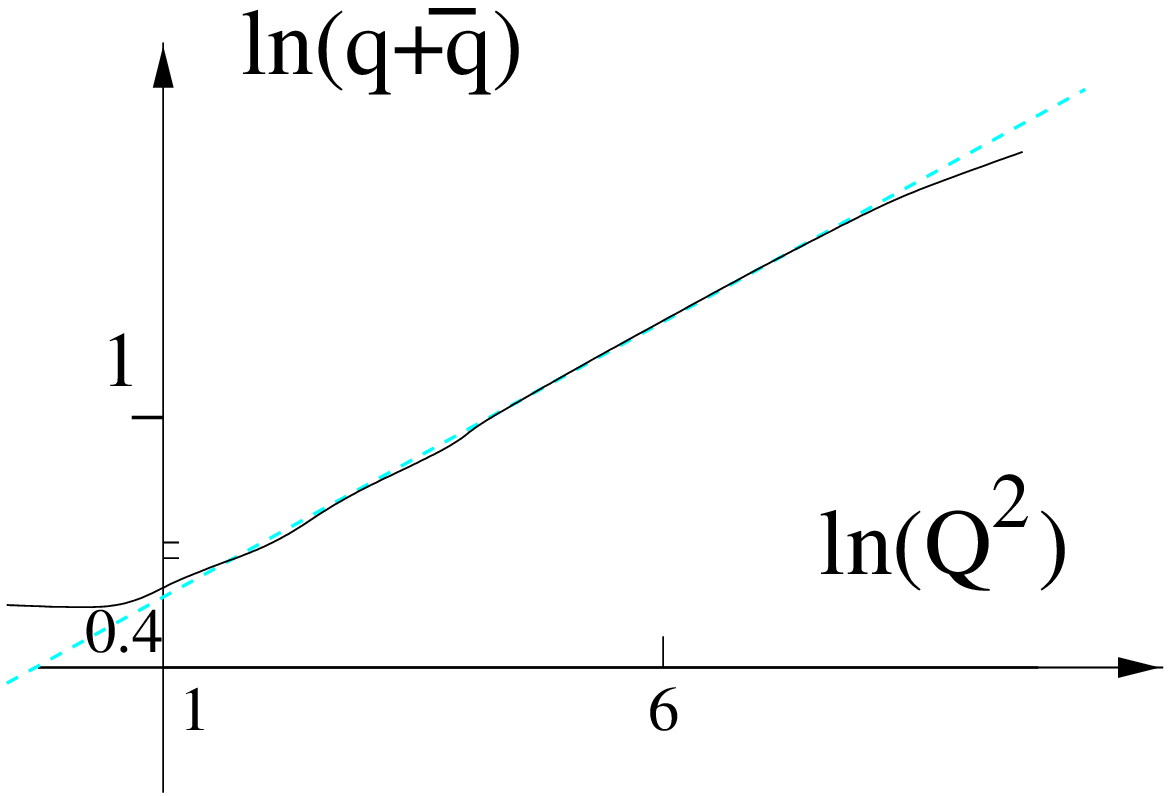}
\includegraphics[width=5cm]{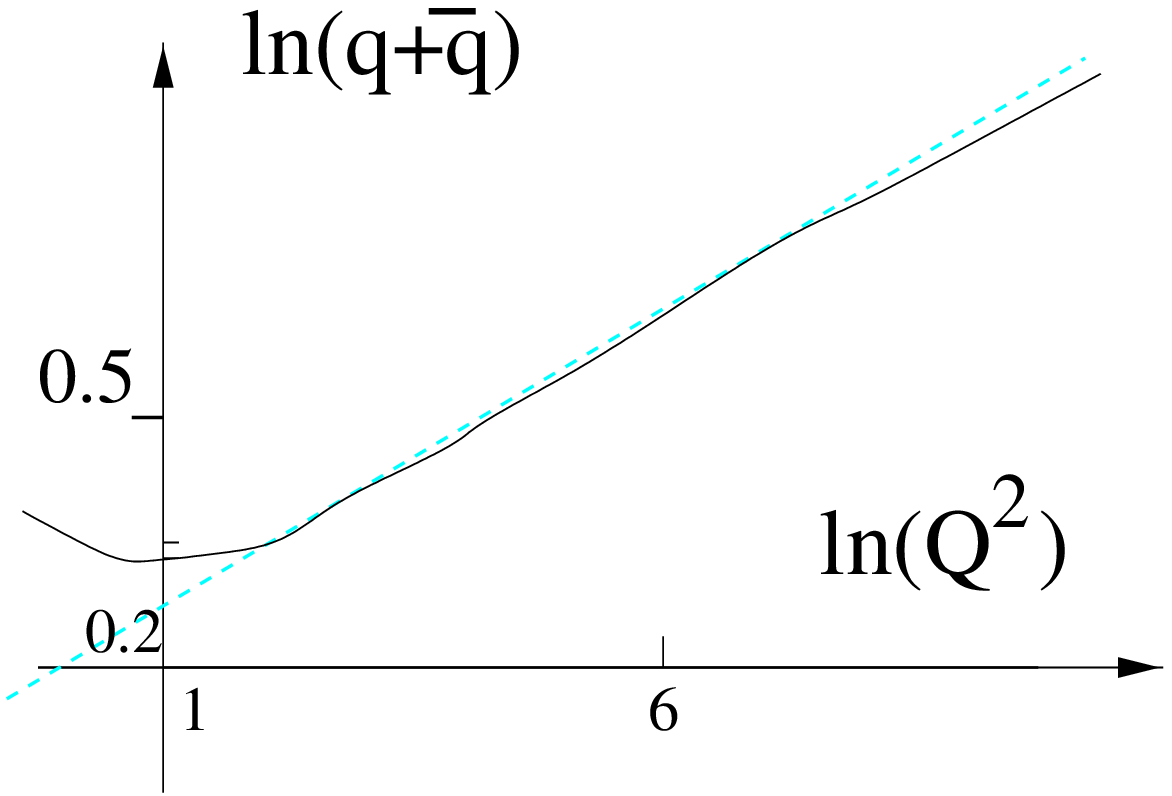} \hspace {2cm}
\includegraphics[width=5cm]{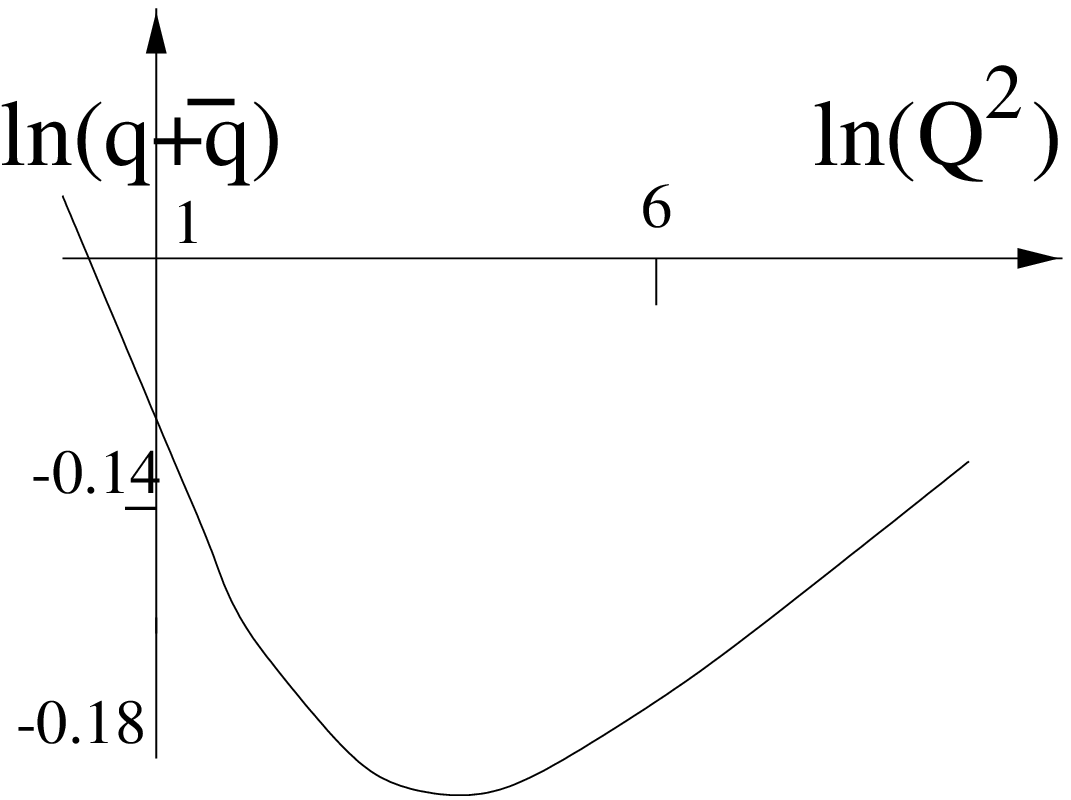}
\caption[]{\it Evolution of the Mellin transformed ${(q+\bar{q})}(j)$ as a 
function of $Q^2.$ \newline {\rm From left to right and top to bottom: 
$j=1.1,1.3,1.4,1.5,1.6,1.8.$}}
\end{center}
\end{figure}

Let us first consider the singlet density 
distribution in Mellin space $\tilde \Sigma \equiv ({q\!+\!\bar q})(j),$ see 
Fig.1. It is an easy exercise to obtain it  from the input GRV parametrizations 
\cite{grv} and LO DGLAP matrix elements in Mellin space \cite{dglap}.

It is clear from Fig.1 that there  exists an interval $1.3 \le j \le 1.7$ 
in which the slope of $\ln \tilde \Sigma$ as a function of $\ln Q^2$ is almost 
constant.   The observed approximate constancy 
meets the requirement\footnote{The NLO 
condition (\ref{i+)}) implies a smooth variation of the slope. Presently, we 
will study the $Q^2$ average only, delaying a more refined (but deserved) study 
of the $Q^2$ dependence of the slope
for a forthcoming publication \cite{us}.} of  condition {\bf i)}. We will 
focus our study to this  region which is 
Mellin-conjugated to the small $x_{Bj}$ region.

Taking into account the anomalous dimension values\footnote{In this preliminary 
study we considered only discrete values $\g_i(j_i),$ where 
$j_i=1.3,1.4,1.5,1.6,1.7.$ determined for a fixed range $2\le \log Q^2 \le 6.$}, 
it is now straightforward to look for the LO prediction (\ref{ii)}) 
and the various NLO predictions  (\ref{ii+)}) depending on the choice of 
resummation scheme in \cite{salam}.

We have displayed in Fig.2 for comparison the results for both the standard (LO) 
BFKL  and for one of the Resummed NLO BFKL schemes ({\it scheme 4}). In our 
application 
plotting $\chi(\g_i(j_i))$ as a function of $ j_i$ should give points aligned 
on  a straight line extrapolating to 0 at $j =1.$ The slope gives the (average) 
value of $\pi N_C/\al_s.$
By simple inspection of Fig.2, it is clear that there is a large difference 
between LO and NLO ({\it scheme 4} of \cite{salam}) results. The LO test 
completely  fails 
in shape and magnitude, while the NLO test is  satisfactory. In this case, the  
measured slope leads to an average value $\al \sim 
.23$ which  is reasonable for the $Q^2$ range considered for the structure 
functions.

\begin{figure}[ht]
\begin{center}
\includegraphics[width=9.5cm]{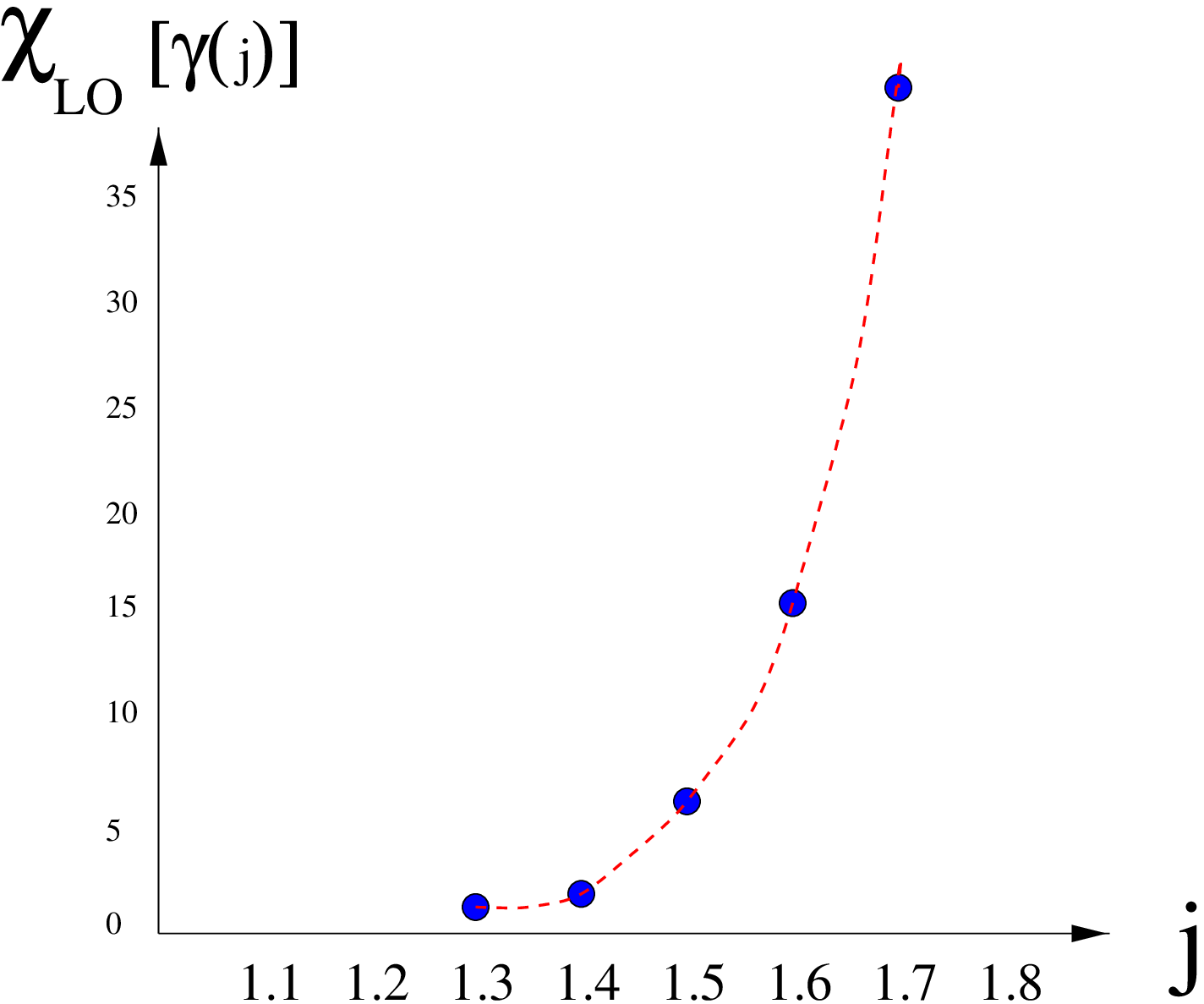} \hspace {2cm}
\includegraphics[width=9.5cm]{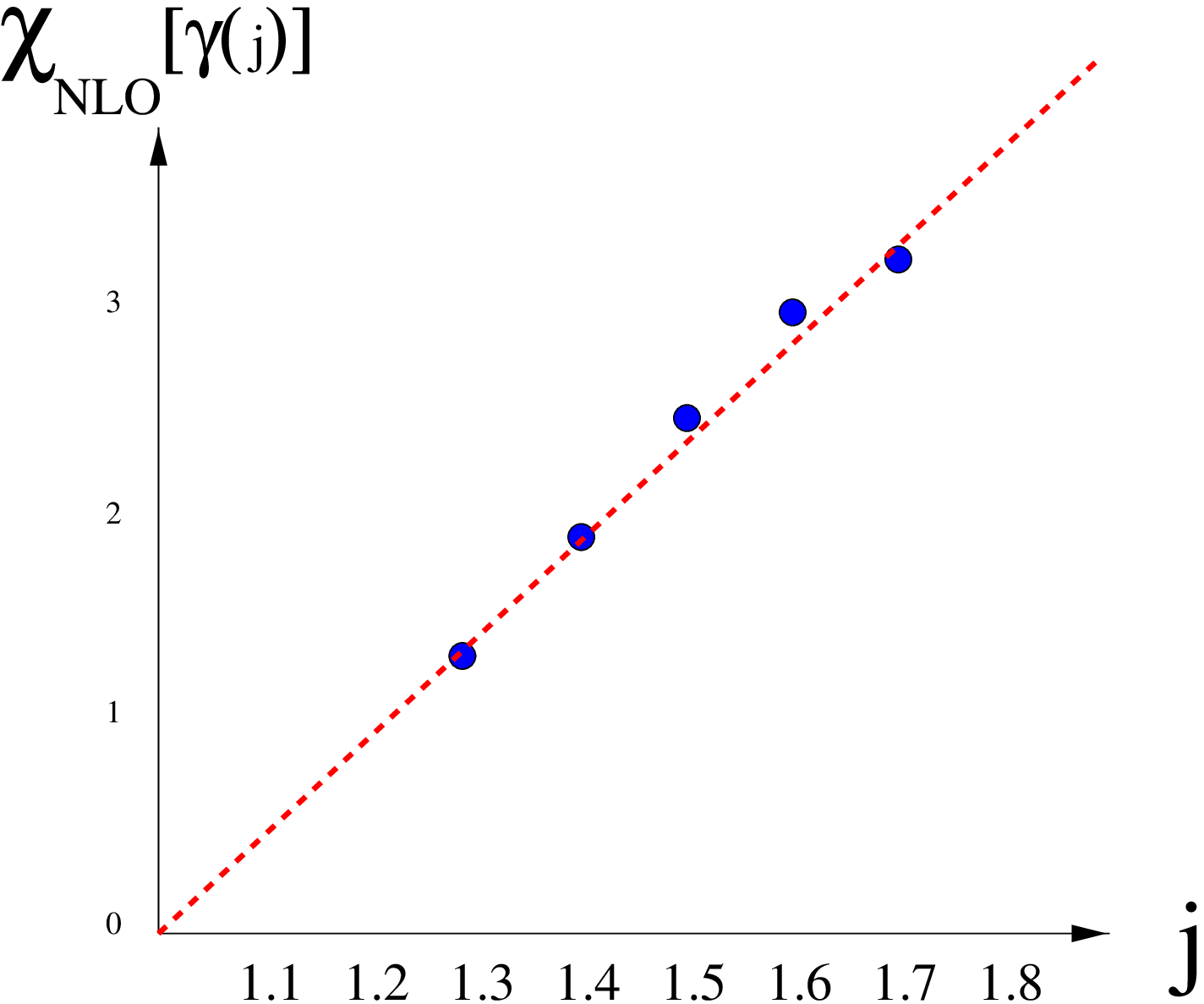}
\caption[]{\it Study of the LO and NLO BFKL kernels in Mellin $j$-space.\newline 
{\rm Top: (LO) BFKL kernel;\newline Bottom: Resummed NLO BFKL kernel, {\it 
scheme 4} 
of Ref. \cite{salam}.}}
\end{center}
\label{en}
\end{figure}
Some comments are in order. We have here used a set of structure functions 
which, on the one hand verifies the DGLAP evolution and, in the other hand give 
a satisfactory fit of data. The conclusion of our test would be that there is 
not much difference between DGLAP evolved {\it effective} anomalous dimensions 
from  DGLAP evolution and from some well-choosen resummed NLO-BFKL kernels. This 
remark 
corroborates the proximity of DGLAP and NLO-BFKL predictions for  the gluon 
anomalous dimension in 
\cite{autres} and, in a different context, the smallness of BFKL-like 
corrections to the DGLAP evolution equation found in \cite{forte} for HERA data. 

The new point is that our method  allows for sensitive tests of the resummed 
NLO-BFKL 
kernels. For instance the  {\it schemes 1,2} fail 
\cite{salomez'} and this is to be related to the fact that not all large logs 
are 
resummed \cite{salam}.  {\it Scheme 3} gives a satisfactory shape but with an 
averaged $\al_s \sim .27$ quite too strong. This is for illustration of the 
sensitiveness of the method.

We postpone to a more systematic study 
\cite{us} the comparison between different sets of structure functions, 
different methods, and also the consideration of non-DGLAP parametrizations in 
order to evaluate the systematic biases which could occur from these different 
options. We also leave for future study the relations (\ref{iii)}),(\ref{iii+)})
which gave  promising results in \cite{salomez}.

\section{Conclusion and outlook}

\begin{itemize}

\item We have proposed a method for confronting with precision ``data'' the 
various resummed BFKL kernels with next-to-leading log accuracy.  These ``data'' 
are the Mellin-transformed of the proton structure function. 

\item Due to the high precision of modern experimental data on $F_2$, we expect 
the 
 Mellin transform to be well determined, at least in the region of $j$ and $Q^2$ 
needed 
for the test. 

\item We make use of a $k_T$ factorized formulation of the structure function 
which grasps the constraints coming from the QCD renormalization group improved 
small-$x_{Bj}$ equation \cite{autres}. 

\item In a first, preliminary, application using the 
parametrization \cite{grv} as input, the method leads with high sensibility  to 
an incompatibility of the LO BFKL kernel while only one of the four resummed NLO 
BFKL schemes of Ref. \cite{salam} shows a compatibility with the DGLAP 
parametrization of data. 

\end{itemize}

The application we have performed is far from complete, and was only serving as 
a test of the method sensitivity. Further studies in various interesting 
directions deserve to be pursued \cite{us}. let us quote some of them.

It is interesting to test whether the present set of data and with which 
accuracy, the Mellin transformed of the structure functions can be obtained. In 
particular, it would be useful to see eventual differences between DGLAP and 
non-DGLAP parametrizations of data, to see whether and where there could be a 
systematical bias introduced by the DGLAP framework.

Concerning the resummation schemes, the application of the method with the 
correct
NLO accuracy requires to take fully into account the coupling constant running. 
It is thus required to separate data in small regions of $Q^2$ and make the 
tests separately in each region, in order to follow the $Q^2$ evolution 
of the coupling constant.

We expect to be able to answer these questions soon \cite{us}.
\section*{Acknowledgement}
The materials for the phenomenological application comes from Julien Salomez 
\cite{salomez,salomez'}, whom I thank. I thank Gavin Salam for fruitful 
discussions and  Christophe Royon and 
Laurent Schoeffel for a fruitful collaboration. 

As well as many of my colleagues, I am grateful to Jan (Kwiecinski) for his 
enthousiasm and readiness of mind, and for his work which represents a 
cornerstone  in our domain. {\it Bonne et Longue Continuation}, Dear Jan!


\begin{thebibliography}{99}

\bibitem{bfkl}
L.N.Lipatov, {\it Sov. J. Nucl. Phys.} {\bf 23} (1976) 642;
V.S.Fadin, E.A.Kuraev and L.N.Lipatov, {\it Phys. lett.} {\bf B60} (1975)
50;
E.A.Kuraev, L.N.Lipatov and V.S.Fadin, {\it Sov.Phys.JETP} {\bf 44} (1976) 45, 
{\bf 45} (1977) 199; 
I.I.Balitsky and L.N.Lipatov, {\it Sov.J.Nucl.Phys.} {\bf 28} (1978) 822.

\bibitem{old} \rr {H Navelet, R.Peschanski, Ch. Royon, S.Wallon} {\it
    Phys. Lett.} {B385} {(1996) 357}. \rr {S.Munier, 
R.Peschanski}{Nucl.Phys.}{B524}{(1998) 377}.
    
\bibitem{next}
V.S. Fadin and L.N. Lipatov, Phys. Lett. B429 (1998) 127; M.Ciafaloni, Phys. 
Lett. 
B429
(1998) 363; M. Ciafaloni and G. Camici, Phys. Lett.  B430 (1998) 349.

\bibitem{salam} \rr{G.P. Salam}{JHEP 9807}{}{(1998) 019}


\bibitem{autres} \rr {M. Ciafaloni, D. Colferai, G.P. 
Salam}{Phys.Rev.}{D60}{114036}, \rr {}{JHEP 9910}{}{(1999) 017};\\
\rr {M. Ciafaloni, D. Colferai, G.P. 
Salam,A.M. Stasto}{Phys.Lett.}{B541}{(2002) 314}.

\bibitem{lipatov}
\rr {Stanley J. Brodsky, Victor S. Fadin, Victor T. Kim, Lev N. Lipatov, 
Grigorii B. Pivovarov}{JETP Lett.}{70}{(1999) 155}.


\bibitem{us} \rr {R.Peschanski, Ch. Royon, L.Schoffel} {} {} {to appear}.

\bibitem{dglap} G.Altarelli and G.Parisi,
{\it Nucl. Phys.} {\bf B126}  18C (1977) 298.
V.N.Gribov and L.N.Lipatov, {\it Sov. Journ. Nucl. Phys.} (1972) 438 and 675.
Yu.L.Dokshitzer, {\it Sov. Phys. JETP.} {\bf 46} (1977) 641. For a review, see 
{\it e.g.} 
\rr{Yu.L.~Dokshitzer, V.A.~Khoze, A.H.~Mueller, S.I.~Troyan} {Basics of 
Perturbative QCD}{}.




\bibitem{kt}
\rr {S. Catani, M. Ciafaloni, F. Hautmann}{Nucl.Phys.}{B366}{(1991) 135};
\rr {S. Collins, R.K. Ellis}{Nucl.Phys.}{B360}{(1991) 3};
\rr {E.M. Levin, M.G. Ryskin, Yu.M. Shabelski, A.G. 
Shuvaev}{Sov.J.Nucl.Phys.}{B53}{(1991) 657}.

\bibitem{bartels}J. \rr {J. Bartels, D. Colferai, S. Gieseke, A. Kyrieleis} 
{Phys.Rev.} {D66} {094017, and references therein}.

\bibitem{salomez} \rr {J. Salomez} {Le mod\`ele des dip\^oles en QCD 
perturbative} {} {(in French), Saclay preprint T02/147 (2002)}, Diploma Memoir 
for the ``DEA Rhône-Alpin, ENS Lyon, France''. Available at: 
http://www-spht.cea.fr/articles/t02/147/

\bibitem{salomez'} J. Salomez, unpublished notes.


\bibitem{grv}\rr {M.  Gluck, E. Reya, A. Vogt} {Z.Phys.} {C67} {(1995) 433}; 
See  
\rr 
{M.  Gluck, E. Reya, A. Vogt} {Eur.Phys.J.} {C5} {(1998) 461}, for updated 
parametrizations.

\bibitem{forte} \rr {R.S. Thorne} {Phys.Rev.} {D60} {(1999) 054031};\rr {G. 
Altarelli, R.D. Ball, S. Forte}{Nucl.Phys}{B621}{(2002) 
359}, and references from the same authors therein.

\end{thebibliography}
\end{document}